\documentclass[12pt]{yoshiken}

\usepackage{amsmath,amssymb,graphicx}

\newcommand{\KUCPlogo}{\hbox{\lower 1.4ex\hbox{\Huge\boldmath $\cal K$}
\kern -1.15em {\sffamily \bfseries\large\ UCP}
\put(-27.5,-6){\tiny\it preprint}}}

\numberwithin{equation}{section}

\begin{document}
\begin{flushright}

\parbox{3.2cm}{
{KUCP-0150\hfill \\
{\tt hep-th/004184}\\
April 2000}
 }
\end{flushright}

\vspace*{3cm}

\begin{center}
 \Large\bf The Effect of the Boundary Conditions  
 in the Reformulation of ${\rm QCD_{4}}$\footnote{ To be published in
 {\bf JHEP}.\\
\hfill \KUCPlogo}
\end{center}

\vspace*{1cm}

\centerline{\large Kentaroh Yoshida}

\begin{center}
{\it Graduate School of Human and Environmental Studies,
\\ Kyoto University, Kyoto 606-8501, Japan. }
\end{center}

\centerline{\tt E-mail:~yoshida@phys.h.kyoto-u.ac.jp}

\vspace*{2.5cm}

\centerline{\bf Abstract}

We discuss the phase structure of ${\rm QCD}_{4}$ as a perturbative deformation
 of the Topological Quantum Field Theory (TQFT). When we choose a
 special Maximal Abelian gauge (MAG) as the
 gauge fixing, the TQFT sector is equivalent
 to a 2D $O(3)$ non-linear sigma model (NLSM).  We consider the finite
 temperature case, and investigate the effect of the boundary conditions  
and the phase 
 structure of the TQFT sector. It can have a deconfining 
 phase under the twisted boundary conditions. However, this phase is screened once pertubative
 parts 
 are added.  We conclude 
 that the information about the phase structure 
 is encoded in the $U(1)$ background in the case of the MAG.

\thispagestyle{empty}
\setcounter{page}{0}

\newpage

\section{Introduction}
It is a long-standing problem to prove the quark confinement. There are some scenarios to explain this phenomenon qualitatively, one of which
is widely known as a dual QCD vacuum scenario\cite{N,M}. However, these are not
sufficient proofs and we need further efforts to investigate this phenomenon.   

Recently  several
authors\cite{Kondo,HT,Izawa} have proposed a novel reformulation of QCD
as a perturbative deformation of a TQFT. By the use of this reformulation, 
a confining-deconfining phase transition has been investigated\cite{HT}.
This work is based on the Kugo-Ojima (KO)
confinement criterion\cite{KO} and  the gauge fixing is performed in
an  $OSp(4|2)$ type Feynman
gauge\cite{OSp}. Then,  the TQFT sector becomes a 2D chiral model through the
Parisi-Sourlas (PS)
mechanism\cite{PS}, equivalently a 2D $O(4)$ NLSM. When it is  extended to the finite
temperature in the real-time formalism\cite{NS}, it has
a deconfining phase under
the twisted boundary conditions through the spontaneous symmetry
breaking (SSB) of an $O(4)$ symmetry but not under the periodic
ones. This result has been showed by the calculation of the effective
potential. In general,
 the  SSB 
in 2D systems
is forbidden by the Coleman-Mermin-Wagner's theorem\cite{MG}. This theorem is
based on the infrared divergence peculiar in 2D systems.  
But, if 
 the twisted boundary conditions are imposed, this divergence is softened and
  the phase transition can take place by the SSB. Therefore, a deconfining
  (massless) phase can appear.  This phase is retained
 as if
pertubative parts are added although the  boundary conditions are
slightly modified. 

The same analysis can be adopted to the case of the $OSp(4|2)$ type MAG\cite{MAG}. 
There, the role of the boundary conditions has not been
clarified. The purpose of our study is to make it clear.
In this case, the TQFT sector becomes a 2D coset model, equivalently a
2D $O(3)$ NLSM, and  we can reach
the similar conclusion about the phase structure in the TQFT sector.
But the phase structure of the TQFT sector is
not retained and only a deconfining phase can survive when
 the perturbative parts are added. The effect of these parts replaces the twisting factors 
in the twisted boundary conditions
 by the unit element and so the twisted boundary 
conditions 
become equivalent to the periodic ones.  
Moreover, we can show that the linear potential remains in the full
${\rm QCD_{4}}$ 
if we 
assume the
$U(1)$ Abelian dominance. This linear  potential  means the quark
confinement in the Wilson criterion. Then, a  Polyakov loops'
correlator decays exponentially at large distance. This result also implies a
confining phase. 
 It may seem
inconsistent to the case of the Feynman type gauge because the only difference is 
the gauge fixing.  
However, we notice that we have not considered the
$U(1)$ background. It turns out that the information about the
phase structure is encoded in it.
 
This paper is organized as follows. In section 2, the
reformulation of the ${\rm QCD_{4}}$ at finite temperature is
introduced. In 
section 3, we 
discuss the effect of the boundary conditions in the TQFT sector and 
investigate how it is
modified when the perturbative parts are added. In section 4, we comment
on the phase structure of the $U(1)$ background. This argument is based
on the work\cite{Kondo2} and the topological object plays an
important role. Finally, in section 5, we explain our results and
discuss 
future problems.

\section{Reformulation of ${\rm QCD_{4}}$ at Finite Temperature}

We start with a finite temperature partition function
\begin{equation}
 Z_{\rm FT} = \int_{\rm periodic} [dA_{\mu}][dC][d\bar{C}][dB]\exp \bigg\{ \frac{1}{2g_{\rm YM}^{2}} {\rm Tr}_{G}F^{2}_{\mu\nu}(A) - i\delta_{\rm B}G_{\rm GF+FP}[A_{\mu},C,\bar{C},
B ] \bigg\}
\end{equation}
where $F_{\mu\nu}\overset{\rm def}{=} \partial_{\mu}A_{\nu} - \partial_{\nu}A_{\mu} -
i[A_{\mu},A_{\nu}]$ and $\delta_{\rm B}$ induces the BRST transformation
\begin{eqnarray}
 \delta_{\rm B}A_{\mu} &=& D_{\mu}[A]C,~~~~\delta_{\rm B}C = iC^{2},  \\
 \delta_{\rm B}\bar{C}\ &=& iB, ~~~~~~~~~~\delta_{\rm B}B = 0. \nonumber 
\end{eqnarray}
The $C(\bar{C})$ is a (anti-)ghost of the system and the $B$ is a
Nakanishi-Lautrup field.
We consider the case of the gauge group
$G=SU(N)$ without quark fields. 
The gauge field $A_{\mu}$ is expressed in terms of  new fields $V_{\mu}$ 
and $U$ 
\begin{equation}
 A_{\mu} = UV_{\mu}U^{\dagger} + \frac{i}{g_{\rm YM}}U\partial_{\mu}U^{\dagger}.
\end{equation}
We shall use the Faddeev-Popov (FP) trick and insert a unit in the path 
integral
\begin{equation}
\label{unit}
 1 = \Delta[A]\frac{1}{N}\sum_{k=0}^{N-1}\int_{B_{k}}[dU]\prod_{x}\delta\big(\partial^{\mu}A^{U^{-1}}_{\mu}(x) \big)
\end{equation} 
where  the $\Delta[A]$ is the FP determinant and the $B_{k}$ denotes the boundary
condition of the  $U(x)$
\begin{equation}
 B_{k}:U(-i\beta,{\bf x}) = U(0,{\bf x})e^{ik\pi/N}~~({\rm for}~ k=0,\cdots,N-1).  
\end{equation}
This condition is fully discussed  in the next section.
Eq.(\ref{unit}) can be rewritten as 
\begin{equation}
 1 = \frac{1}{N}\sum_{k=0}^{N-1}\int_{B_{k}}[dU]\int_{\rm periodic}[d\gamma][d\bar{\gamma}][d\beta]\exp\Big\{i\int d^{4}x[-i\tilde{\delta}_{\rm B}\tilde{G}_{\rm GF+FP}(V_{\mu},\gamma,\bar{\gamma},\beta)]\Big\}
\end{equation}
where the new BRST transformation $\tilde{\delta}_{\rm B}$ acts on the
fields $V_{\mu},~\gamma,~\bar{\gamma}$ and $\beta$ as 
\begin{eqnarray}
 \tilde{\delta}_{\rm B}V_{\mu} &=& D_{\mu}[V]\gamma,~~ 
 \tilde{\delta}_{\rm B}\gamma = i\gamma^{2}, \nonumber \\
 \tilde{\delta}_{\rm B}\bar{\gamma} &=& i\beta,~~~~~~~~\tilde{\delta}_{\rm B}\beta = 0.
\end{eqnarray} 
Here we used a formula of the $\tilde{G}_{\rm GF + FP}$ 
\begin{equation}
 \tilde{G}_{\rm GF+FP} \overset{\rm def}{=} {\rm Tr}_{G}(\bar{\gamma}\partial^{\mu}V_{\mu}).
\end{equation}
Thus, we can obtain the following partition function
\begin{eqnarray}
 Z_{\rm FT} & =& \frac{1}{N}\sum_{k=0}^{N-1}\int_{B_{k}}[dU]\int_{\rm periodic}[dV_{\mu}][dC][d\bar{C}][dB][d\gamma][d\bar{\gamma}][d\beta]  \nonumber \\
& &\exp\Big\{i\int d^{4}x \Big[ \frac{1}{2g_{\rm YM}}{\rm Tr}_{G}F_{\mu\nu}^{2}(V) - i\tilde{\delta}_{\rm B}\tilde{G}_{\rm GF+FP}[V_{\mu},\gamma,\bar{\gamma},\beta]  \nonumber \\
& & -i\delta_{\rm B}G_{\rm GF+FP}\big[\frac{i}{g_{\rm YM}}U\partial_{\mu}U^{\dagger} + UV_{\mu}U^{\dagger},C,\bar{C},B\big]          \Big]           \Big\}.  
\end{eqnarray}
Next, let us specify the gauge fixing term $G_{\rm GF+FP}$. In the work\cite{Kondo},
 an $OSp(4|2)$ type MAG is used
\begin{equation}
 G_{\rm GF+FP} = \bar{\delta}_{\rm B}{\rm Tr}_{G/H}[A_{\mu}^{2} + 2i C\bar{C}].
\end{equation}
The $H$ is the maximal Abelian subgroup of the G.
On the other hand, an  $OSp(4|2)$ type Feynman gauge is utilized in the work\cite{HT}
\begin{equation}
 G_{\rm GF+FP} = \bar{\delta}_{\rm B}{\rm Tr}_{G}[A_{\mu}^{2} + 2i C\bar{C}].
\end{equation}
These gauge fixing conditions lead to the following 2D  TQFT sectors
through the PS mechanism, respectively
\begin{eqnarray}
\label{coset}
 S_{\rm TQFT} &=&  \frac{\pi}{g_{\rm YM}^{2}} \int d^{2}x {\rm Tr}_{G/H}[\partial_{\mu}U\partial_{\mu}U^{\dagger}]~~~~{\rm (Coset~ model)},                   \\
S_{\rm TQFT} &=& \frac{\pi}{g_{\rm YM}^{2}} \int d^{2}x {\rm Tr}_{G}[\partial_{\mu}U\partial_{\mu}U^{\dagger}]~~~~{\rm (Chiral~ model)}.
\label{chiral}
\end{eqnarray}
It should be noted that the difference of the two models is associated
with the 
degrees of freedom in the maximal torus part $H$.
 In particular, the weak coupling limit ($g_{\rm YM} \rightarrow 0$) 
of the finite temperature ${\rm
QCD_{4}}$ is described by the TQFT sector with summing over all the boundary conditions
\begin{equation}
 Z_{\rm FTTQFT} = \frac{1}{N}\sum_{k=0}^{N-1}\int_{B_{k}}[dU]\int_{\rm periodic}[dC][d\bar{C}][dB]\exp \{ i S_{\rm TQFT} \}.
\end{equation}

\section{ Study of Boundary Conditions}

We restrict ourselves to the case of $G=SU(2)$  for simplicity. 
The real-time and imaginary-time formalisms\cite{NS,TFD,RI} are standard methods to
deal with the finite temperature system. Both formalisms extend the time
coordinate to a complex time and the fields obey the (anti-)periodic
boundary conditions. In particular, the gauge field obeys a periodic condition
\begin{equation}
  A_{\mu}(-i\beta,{\bf x}) = A_{\mu}(0,{\bf x}).
\end{equation}
The gauge field in the TQFT sector has a pure gauge type configuration,
and 
the twisted boundary
conditions are allowed:
\begin{equation}
\label{twist}
 B_{g}: U(-i\beta,{\bf x}) = U(0,{\bf x})g ~~({\rm for}~\forall~ g \in~{\rm global}~ SU(2)).
\end{equation}  
Let us consider the TQFT sectors in eqs.(\ref{coset}), (\ref{chiral}). 
 It is proved that the
TQFT sector (\ref{chiral}), equivalently an $O(4)$ NLSM,  has a
deconfining phase 
under the twisted boundary conditions ($g \neq {\bf 1}$) in the real-time formalism\cite{HT}.
 But when one imposes usual periodic boundary conditions ($g = {\bf 1}$),
this phase does not appear. Also, this phase transition cannot take
 place in the imaginary-time formalism. This point is discussed later
 in this section.
 
Next, we investigate the phase structure of the TQFT
sector (\ref{coset}). Our main purpose is to study it.  
The coset model action (\ref{coset}) 
can be rewritten as\footnote{We omit the ghost term.}
\begin{equation}
 S_{\rm TQFT} = \frac{\pi}{g_{\rm YM}^{2}}\int d^{2}x [\partial_{\mu}{\bf n}\cdot\partial_{\mu}{\bf n}],  ~~~~~{\bf n}\cdot{\bf n} = 1.
\end{equation}
This is the action of an $O(3)$ NLSM.
Here we used the Euler angle representation of the $SU(2)$ matrix 
\begin{eqnarray}
 U(x) &=& \exp\{i\chi(x)\sigma_{3}/2\}\exp\{i\theta(x)\sigma_{2}/2\}\exp\{i\varphi(x)\sigma_{3}/2\} \nonumber \\
      &=& \left(
\begin{array}{cc}
\exp\{\frac{i}{2}(\varphi +\chi )\}\cos\frac{\theta }{2} & \exp\{-\frac{i}{2}(\varphi -\chi )\}\sin\frac{\theta }{2} \\
-\exp\{\frac{i}{2}(\varphi -\chi )\}\sin\frac{\theta }{2} & \exp\{-\frac{i}{2}(\varphi +\chi )\}\cos\frac{\theta }{2}
\end{array}
\right)
\end{eqnarray} 
and  we parameterize the unit vector field ${\bf n}(x)$ (${\bf  n}:~{\bf
R}^{2} \longrightarrow S^{2}$) as
\begin{equation}
 {\bf  n} \overset{\rm def}{=} 
\begin{pmatrix}
 n^{1}(x) \\
 n^{2}(x) \\
 n^{3}(x)
\end{pmatrix}
\overset{\rm def}{=} 
\begin{pmatrix} 
\sin\theta(x)\cos\varphi(x) \\
\sin\theta(x)\sin\varphi(x) \\
\cos\theta(x)
\end{pmatrix}
.
\end{equation}
Then, we have to determine the boundary condition on the ${\bf n}(x)$.
Useful relations can be used in the calculation\footnote{We normalize the
generators $T^{A}$'s of the SU(2) as
${\rm Tr}_{G}(T^{A}T^{B})= \frac{1}{2}\delta^{AB}$.} :
\begin{equation}
 n^{A}(x)T^{A}=U^{\dagger}(x)T^{3}U(x),~~n^{A}(x)=2{\rm Tr}_{G}[U(x)T^{A}U^{\dagger}(x)T^{3}]~~(A=1,2,3).
\end{equation}
We find that the ${\bf n}(x)$ is invariant under the  $U(1)$ transformation
generated by the $T^{3}$ and it   can be rotated by generators
associated with the coset $SU(2)/U(1)$.
Also, eq.(\ref{coset}) has the following global $SU(2)_{\rm L}$ symmetry
\begin{equation}
 U \longrightarrow Uh, ~~~h\in SU(2)_{\rm L}.
\end{equation} 
Then  the ${\bf n}(x)$  transforms as 
\begin{equation}
 n^{A}T^{A} \longrightarrow n^{A}(h^{\dagger}T^{A}h) \overset{\rm def}{=} {n'}^{A}T^{A}
\end{equation}
and we easily find an action of the element $h$ on $n^{A}$
\begin{equation}
\label{n}
 n^{A} \longrightarrow {n'}^{A} = \sum_{B=1}^{3} {\rm ad}(h^{\dagger})^{A}_{~B}n^{B}.
\end{equation}
This means that the {\bf n}$(x)$ is transformed under the $SO(3)$ rotation
but it  is invariant by an action of  the center ${\bf Z}_{2}$ of the $SU(2)$.
By the use of eq.(\ref{n}), the boundary condition on the $U(x)$
can be  
translated into
that on the field {\bf n}$(x)$
\begin{equation}
 n^{A}(-i\beta,x_{1}) = \sum_{B=1}^{3}{\rm ad}(g^{\dagger})^{A}_{~B}n^{B}(0,x_{1}),
\end{equation}
where $x_{1}$ is a spatial coordinate. 
 We can calculate the effective potential under this condition and 
show that 
 the $O(3)$ NLSM has a
deconfining (massless) phase in the real-time formalism under the twisted boundary
conditions (ad($g^{\dagger}$)$\neq {\bf 1}$).
On the other hand, under the periodic boundary conditions (ad$(g^{\dagger})={\bf
1}$),
  it has not this phase as in
 the case of the Feynman type gauge. Also, no massless phase appears in imaginary-time formalism.

The above discussion is limited to the TQFT sector. That is, the gauge
field configuration is the pure gauge one. What happens when one
incorporates 
the gauge adjoint part (i.e.the perturbative part)? This part replaces
the twisted boundary conditions (\ref{twist}) by the following ones
\begin{equation}
\label{center}
 B_{k}: U(-i\beta,{\bf x}) = U(0,{\bf x}){e^{i k \pi/2}} ~~(k=0,~1). 
\end{equation}
Then, the ${\bf n}(x)$ obeys the periodic boundary conditions:
\begin{equation}
 B_{k}: {\bf n}(-i\beta,x_{1}) = {\bf n}(0,x_{1}) ~~(k=0,1).
\end{equation}
Therefore,  all the TQFT sectors have the periodic boundary conditions
\begin{eqnarray}
\frac{1}{N}\sum_{k=0}^{N-1}\int_{B_{k}}[dU] &=& \frac{1}{N}\sum_{k=0}^{N-1}\int_{B_{k}}[dn]
\longrightarrow\int_{\rm periodic}[dn]~~~~(N=2),  \\
& &\frac{1}{N}({\rm periodic }~ +~ {\rm twisted}~ +~\cdots )\longrightarrow {\rm periodic}. \nonumber
\end{eqnarray}
That is, the sum over the boundary conditions becomes an integral 
of the periodic field ${\bf n}(x)$. Because a massless phase cannot exist
under the periodic conditions, only a confining phase
can exist. Once the gauge adjoint part is added, this phase
 of the TQFT sector is screened. We  conclude that the phase
structure of the TQFT sector is hidden in the full ${\rm QCD_{4}}$ with the MAG.

Moreover, if we assume the Abelian dominance, we can interpret this
result in terms of the Polyakov loop. Then, 
we can see that  the correlator of Polyakov loops $P$ and $P^{\dagger}$ is equivalent to
the expectation value of an Abelian  Wilson loop $W$ in the TQFT sector
as shown in Figure 1
\begin{equation}
\label{poly}
 \langle P^{\dagger}({\bf x})P(0)\rangle_{\rm TQFT} \overset{\rm def}{=} \langle W \rangle_{\rm TQFT} = \exp (- \beta V).
\end{equation} 
Here the $V$ is a static potential or a free energy. We can reconstruct
 the expectation value in the full ${\rm QCD_{4}}$ $\langle
 P^{\dagger}({\bf x})P(0)\rangle$
 from eq.(\ref{poly}). Then, the Coulomb potential is added to the V.

 In the $|{\bf x}|\rightarrow \infty$ limit, we obtain a decomposition
\begin{equation}
 \langle P^{\dagger}({\bf x})P(0)\rangle \longrightarrow |\langle P \rangle |^{2}.
\end{equation} 
From this formula, we conclude that if $\langle P \rangle = 0$, the free
energy blows up for the  large separation of the
quarks($|{\bf x}|$$\rightarrow \infty$). We  can interpret this result as a
signal of the 
confinement:
\begin{equation}
 \langle P \rangle = 0,~~({\rm confinement}).
\end{equation}
On the other hand, if $\langle P \rangle \neq 0$, then the free energy
of a static quark-antiquark pair approaches a constant for the limit
$|{\bf x}|$$\rightarrow \infty$. We can interpret this as an evidence of the deconfinement:
\begin{equation}
 \langle P \rangle \neq 0,~~({\rm deconfinement}).
\end{equation}
The TQFT sector in the MAG, $O(3)$ NLSM has instanton solutions and  
it can be shown that the instanton gas induces the linear 
potential in the 
$V$\cite{Kondo}. So  we obtain the $\langle P \rangle =0$  and this
result implies the confinement in the full ${\rm QCD_{4}}$.\footnote{ 
In
this case, we would have to use the imaginary-time formalism.}
 
Thus, we find that only a confining phase can appear in the analysis of
 the TQFT sector and so this point is
 different from the case of the Feynman type gauge.
  We will explain this point in the next section.

Here, we discuss the two methods to deal with the finite temperature
system: the real-time formalism and the imaginary-time one. 
It is amazing that these methods lead to  different results as commented in the work\cite{HT}
when they are applied to the same system. That is, the
phase transition  occurs under the twisted boundary conditions in the  
real-time formalism but not in the imaginary-time 
one.
The  KO confinement criterion is only applicable to the system with a 
continuous
4-momentum. Otherwise the imaginary-time formalism can not be applied and we
have no 
contradiction. 
We may suspect to confront an issue in the MAG case. 
However, it would not so. It is 
 the reason that the physical parts (gauge adjoint parts) really exist
 in the real world and only a confining phase can appear. Therefore, 
 we 
cannot observe this difference of the two methods.  We conclude that the
difference of the two methods is not relevant to the phase structure of the  ${\rm QCD}_{4}$ in the MAG case.

\begin{figure}
 \begin{center}
\includegraphics[width=13cm]{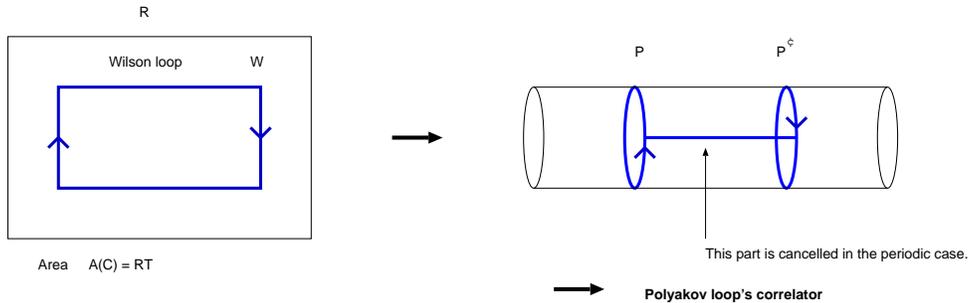}
\caption{Polyakov loops' correlator. ~~{\footnotesize The expectation
  value of a Wilson loop $W$ is equivalent to a correlator of Polyakov
  loops $P$ and $P^{\dagger}$
  in the case of the periodic boundary conditions.}}
 \end{center}
\end{figure}

\section{Comments on Gauge Fixing and $U(1)$ Background}

In this section, we comment on the relation between the phase transition
mechanism and the $U(1)$ background. First, let us
recall that we have chosen the $OSp(4|2)$ type MAG. This gauge spoils
 the chiral $SU(2)_{\rm L}\otimes SU(2)_{\rm R}$ symmetry which
exists in the Feynman type gauge. Eventually, only the $SU(2)_{\rm 
L}$ symmetry survives and the $SU(2)_{\rm R}$ symmetry is lost in the MAG
case. Then, we obtain a topological
object ``monopole'' peculiar to the MAG and it is absent in the
Feynman gauge. This is an $U(1)$ Abelian mopole. Because the
chiral symmetry is broken, we would not be  able to 
apply the KO criterion. However, we have the topological object in our
hand.  Monopoles are interpreted as  instantons in the 2D theory reduced
by the PS mechanism in this case. These play an important role in the confinement of the Wilson
criterion\cite{Kondo}. It is also conjectured that these play a central
role in the phase
transition as well. In the above study, we saw that the
phase structure of the TQFT sector was screened and only a confining
phase can appear. However, 
in the case of the MAG, the information about the phase transition mechanism
is  encoded in  the $U(1)$ background and the
 Berezinskii-Kosterlitz-Thouless (BKT) transition\cite{BKT} occurs by the
 condensation of the topological object.
This would be clarified after we  integrate out the non-diagonal gauge
components. That manipulation is based on the $U(1)$ Abelian dominance
and really was done in the work\cite{Kondo3}. Then, we can obtain
an Abelian projected effective theory. This is just the  $U(1)$ theory 
 but is  asymptotically free. When the scheme of the reformulation is
applied to this $U(1)$ theory\cite{Kondo2}, the TQFT sector of it 
becomes a 2D
$O(2)$ NLSM with a vortex solution.
It is widely known that
 a BKT phase 
transition is induced by the condensation of the vortexes  in this
$O(2)$ NLSM. In this theory, by using parameterizations of the fields $U$ and 
${\bf S}$
\begin{equation}
  U(x) = {\rm e}^{i\varphi(x)},~~{\bf S}(x)=(\cos\varphi(x),\sin\varphi(x)),
\end{equation}
we obtain the TQFT sector of the effective $U(1)$ theory:
\begin{equation}
 S_{\rm TQFT} = \frac{\pi}{g^{2}_{\rm YM}}\int d^{2}x \partial_{\mu}{\bf S}(x)\partial_{\mu}{\bf S}(x).
\end{equation}
Also, the ${\bf S}$ obeys the boundary
condition at finite temperature. But we can easily show 
that the boundary conditions on it becomes equivalent to the periodic 
ones when the perturbative parts are added. And so we cannot state the existence of the phase
transition by the SSB in the TQFT
sector as well. But, at least, we can find a BKT type phase
transition by the $U(1)$ vortex condensation from the confining phase to
the deconfining one. Moreover,  
it is
commented in the work\cite{Kondo3} that  the $U(1)$ vortex is closely relevant to the 
instanton in the  $O(3)$ NLSM. This possibly implies that the vortex condensation
induces an instanton condensation and the confining potential  would 
vanish. This would just correspond to the phase transition to the deconfining
phase.

\begin{figure}
 \begin{center}
\includegraphics[width=13cm]{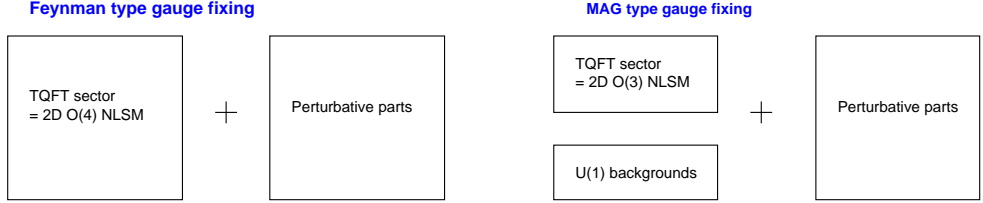}
%  \leavevmode    %kurama ¤Î½ñ¼°
%  \epsfbox{comp.eps} 
\caption{Comparison Feynman type gauge with MAG type gauge in $G=SU(2)$. ~~{\footnotesize The phase structure
  of the TQFT sector is preserved in the case of the Feynman type gauge
  fixing. However, it is screened by the perturbative fluctuation in the 
  case of the MAG type gauge fixing, and its information is encoded in the
  $U(1)$ backgrounds. }}
 \end{center}
\end{figure}

\section{Discussion and Conclusion}

We have investigated the effect of the boundary conditions and found that
it depends 
on the gauge fixings. 
Because the analysis based on the 
KO criterion does not depend on the complicated dynamical
information about the ${\rm QCD_{4}}$, we could not derive the
mechanism of the confinement directly. On the other hand, when we choose the MAG,
we can understand the dynamical mechanism in terms  of monopoles. But
because 
the analysis depends on the special gauge, 
we cannot know how reliable it is. Also, we could not discuss the Higgs
phase. It is an interesting problem to study this phase by the use of
the reformulation of ${\rm QCD_{4}}$ 

  The phase structure of the ${\rm QCD_{4}}$ has investigated in the $OSp(4|2)$ type
MAG and Feynman gauge.
These investigations are based on the toy models where the special
$OSp(4|2)$ type gauge fixing is utilized. But we could explain the results based on the KO 
criterion in the framework of  the Wilson criterion, and 
believe that our considerations  shed light 
on an interrelation  between the works based on the  KO criterion  
and those on the Wilson criterion. 
We hope the paper proceeds the understanding of the quark confinement.

\vspace*{1cm}
\noindent
{\Large \bf Acknowledgements}\\

\noindent
The author would like to thank W.~Souma for  valuable discussions
and K.~Sugiyama for careful reading the manuscript and useful comments.


\begin{thebibliography}{30}

\bibitem{N}Y.~Nambu, Phys. Rev. {\bf D 10}, 4262 (1974).

\bibitem{M}S.~Mandelstam, Phys. Rep. {\bf 23}, 245 (1976).

\bibitem{Kondo}K.-I.~Kondo, 
Phys. Rev. {\bf D 58}, 105019 (1998), {\tt hep-th/9801024}. 

\bibitem{HT}H.~Hata and Y.~Taniguchi, 
	  Prog.~Theor.~Phys.~{\bf 94}, 435 (1995), {\tt hep-th/9502083}; \\
Prog.~Theor.~Phys.~{\bf 93}, 797 (1995), {\tt hep-th/9405145}.

\bibitem{Izawa}K.I.~Izawa, Prog.~Theor.~Phys.~{\bf 90}, 911 (1993).

\bibitem{KO}T.~Kugo and I.~Ojima, Prog. Theor. Phys. Suppl. {\bf 66}, 1
	(1979).

\bibitem{OSp}R.~Delbourgo and P.D.~Jarvis, J.~Phys.~{\bf A 15}, 611 (1982).

\bibitem{PS}G.~Parisi and N.~Sourlas, Phys.~Rev.~Lett.~{\bf 43}, 744 (1979).


\bibitem{NS}A.J.~Niemi and G.W.~Semenoff, Ann. Phys. {\bf 152}, 105 (1984).



\bibitem{MG}N.D.~Mermin and H.~Wagner, Phys.~Rev.~Lett.~{\bf 17}, 1133
	(1966). \\
N.D.~Mermin, J.~Math. Phys. {\bf 8}, 1061 (1967). \\
S.~Coleman, Commun. Math. Phys. {\bf 31}, 259 (1973).



\bibitem{MAG}G.'t Hooft, Nucl. Phys. {\bf B 190}, 455 (1981).

\bibitem{Kondo2}K.-I.~Kondo,
Phys. Rev. {\bf D 58}, 085013 (1998), {\tt hep-th/9803133}.


\bibitem{TFD}Y.~Takahashi and H.~Umezawa, Collective Phenomena {\bf 2},
	55 (1975). \\
H.~Umezawa, H.~Matsumoto and M.~Tachiki, {\it ``Thermo Field Dynamics and
	Condensed States''} (North-Holland, Amsterdam, 1982).
\bibitem{RI}N.P.~Landsman and Ch.G.~van Weert, Phys. Report. {\bf 145},
	141 (1987).

\bibitem{BKT}V.L.~Berezinskii, Soviet Physics JETP {\bf 32}, 493
	(1971). \\
J.M.~Kosterlitz, J.~Phys. {\bf C 7}, 1046 (1974). \\
J.M.~Kosterlitz and D.V.~Thouless, J.~Phys. {\bf C 6}, 1181 (1973).

\bibitem{Kondo3}K.-I.~Kondo, Phys.~Lett.~{\bf B 455}, 251 (1999),
	{\tt hep-th/9810167};~Phys. Rev. {\bf D 57}, 7467 (1998),
	{\tt hep-th/9709109};~Prog. Theor. Phys. Supplement, {\bf 131}, 243
	(1998), {\tt hep-th/9803063}.

\end{thebibliography}
\end{document}